\documentclass[a4paper]{jpconf}
\usepackage{graphicx}
\usepackage{wasysym}
\usepackage{color}
\usepackage{soul}

\newcommand{\sm}{\mathrm{SM}}

\newcommand{\cl}{\mathrm{cl}}

\newcommand{\WS}{\mathrm{WS}}

\newcommand{\sat}{\mathrm{sat}}
\newcommand{\sym}{\mathrm{sym}}

\newcommand{\coul}{\mathrm{Coul}}

\newcommand{\surf}{\mathrm{surf}}
\newcommand{\curv}{\mathrm{curv}}

\begin{document}
\title{Neutron star crust properties: comparison between the compressible liquid-drop model and the extended Thomas-Fermi approach}

\author{G. Grams}
\address{
Institut d’Astronomie et d’Astrophysique, CP-226, Université Libre de Bruxelles, 1050 Brussels, Belgium.}
\ead{guilherme.grams@ulb.be}

\author{J. Margueron and R. Somasundaram}
\address{Univ Lyon, Univ Claude Bernard Lyon 1, CNRS/IN2P3, IP2I Lyon, UMR 5822, F-69622, Villeurbanne, France.}

\author{N. Chamel and S. Goriely}
\address{
Institut d’Astronomie et d’Astrophysique, CP-226, Université Libre de Bruxelles, 1050 Brussels, Belgium.}

\begin{abstract}
We present a detailed analysis of three models predicting the properties of non-uniform matter in the crust of neutron stars: the compressible liquid-drop model, the fourth order Extended Thomas Fermi (ETF) method, and ETF plus Strutinsky integral (ETFSI) correction. The former treats the nuclear clusters as uniform hard spheres, the second takes into account the density distribution which can be different for neutrons and protons, and the last one includes the proton shell effects within the Strutinsky approach. The purpose of this work is to understand the importance of the improvements in the nuclear modeling and to analyze the quantities which are the most sensitive to them. We find that thermodynamic quantities such as pressure, energy and chemical potential, as well as the electron fraction, are in very good agreement among the three models. This confirms previous results where we have shown that the improvement in the finite-size description of the nuclear clusters has a small impact on these quantities, since they are mainly constrained by the bulk properties. The refinements in the finite-size modeling are shown to impact mostly the composition of the nuclear clusters ($Z_\cl$, $N_\cl$) in an ordering which ranks according to the leptodermous expansion. 
This analysis is performed considering both the r-cluster and the e-cluster representations.
The proton shell effects are shown to stabilize $Z_\cl$, which consequently impacts the neutron number $N_\cl$ as well.
\end{abstract}

\section{Introduction}

Neutron stars (NSs) are excellent laboratories to test nuclear physics models under extreme conditions of density and isospin asymmetry~\cite{Rezzolla2018}. Nucleons and nuclear clusters are present in the crust of NSs and nucleons in their core. More specifically NSs core is assumed to be composed by uniform nuclear matter in beta-equilibrium with electrons and leptons, and the crust, where the low-density baryonic fields break the translation symmetry and prefer forming clusters. 
The outer crust contains a mixture of electrons and nuclear clusters, complemented with a neutron fluid in the inner crust. 
In this study, we focus on the properties of the crust where the modeling of the clusters represents an important challenge.
We adopt the compressible liquid-drop model (CLDM) approach, as in Ref.~\cite{Carreau2019a,Grams2022a}, where the bulk term is fixed by the nuclear meta-model (MM)~\cite{Margueron2018a} and the finite-size (FS) term contains Coulomb, surface and curvature contributions as described by the FS4 approximation presented in Ref.~\cite{Grams2022a}.
The FS terms are also fine-tuned on experimental nuclear masses, as suggested by Steiner~\cite{Steiner2008} since they are derived in the CLDM where the smoothing effect of the surface is disregarded. Regarding the uncertainties driven by modelling the NS crust, we however found that the FS terms of the CLDM play a minor role in comparison with the bulk terms. We have analyzed for instance the importance of the symmetry energy, which controls several properties on the NS crust through the $\beta-$equilibrium condition. A clear example being the electron fraction~\cite{Grams2022b}.

While smaller than the contribution of the bulk terms, the FS terms are however  important and could potentially impact transport properties in the crust. We have shown \cite{Grams2021a,Grams2022a,Grams2022b} that the CLDM based on the MM for the bulk properties is a very useful tool to investigate uncertainties in model predictions. In recent papers we have analyzed the impact of different treatment in the FS terms existing in the literature and we have observed that their impact ranks according to the leptodermous expansion of the liquid-drop model~\cite{Grams2022a,Grams2022b}. 
We have observed that the cluster modelling impacts the mass and charge of the nucleus in the crust as well as the crust-core transition density but little effect is seen on cluster isospin, overall proton fraction and volume fraction occupied by the cluster. We have also analyzed predictions guided by several $\chi_{EFT}$ Hamiltonians and we have estimated their ability to reproduce nuclear masses. Interestingly, we found that not all $\chi_{EFT}$ Hamiltonians provide nuclear properties in accord with experimental data. A selection of $\chi_{EFT}$ Hamiltonians based on their goodness in reproducing the experimental masses has then been done \cite{Grams2022a}. For low density neutron matter(NM) existing in the crust of NSs, the different $\chi_{EFT}$ Hamiltonians converge to very similar predictions. 

We have also discussed the impact of the surface parameter $p_\surf$ controlling the surface tension at large isospin fraction \cite{Grams2022a}. This parameter is not constrained by nuclear masses but play an important role in the high density region of the NS crust~\cite{Grams2022a}. In the CLDM it is difficult to adjust precisely the value for this parameter since it plays a minor role for finite nuclei. Specific modeling of very neutron-rich nuclei could be employed, e.g., slab geometries, or direct comparison to more microscopic approaches as the ETF model. In the present work, we show that it is indeed possible to fine-tune the parameter $p_\surf$ to get a very good agreement between the CLDM and the ETF predictions for the cluster neutron number $N_\cl$.

In the following, we will show detailed results comparing the predictions based on three models (CLDM, ETF, ETFSI) based on the same Skyrme interaction, which is taken to be BSk24~\cite{Goriely2013}. In the CLDM approach that we employ, the bulk properties are obtained from the MM~\cite{Margueron2018a} which is adjusted to reproduce the same predictions of the original BSk24 Skyrme interaction. We thus first present how this adjustment is performed and then we  discuss the properties of non-uniform matter in the crust of NSs. The CLDM approach is described in details in the Refs.~\cite{Grams2021a,Grams2022a,Grams2022b} while the ETF and ETFSI are from Ref.~\cite{Pearson2018}.

\section{Uniform matter predictions from BSk24 and comparison to the meta-model}

The predictions for uniform matter based on BSk24 Skyrme interaction have been published in Ref.~\cite{Goriely2013}. Here, we describe how the MM \cite{Margueron2018a,Somasundaram2021} is constructed upon these existing predictions.

The MM is designed to reproduce the topological properties of the energy per particle around saturation density, which are encoded into the nuclear empirical parameters (NEP), e.g. $E_\sat$, $K_\sat$, $E_\sym$, $L_{\sym}$, defined as the coefficients of a Taylor expansion of the binding energy per particle in symmetric matter (SM), $e_\sm(n)$, and the symmetry energy $e_\sym(n)$,
\begin{eqnarray}
e_\sm(n) &=& E_\sat + \frac 1 2 K_\sat x^2 + \frac{1}{6} Q_\sat x^3 
+ \frac{1}{24} Z_\sat x^4+\dots \, ,  \\ 
e_\sym(n) &=& E_\sym + L_\sym x + \frac 1 2 K_\sym x^2 + \frac{1}{6} Q_\sym x^3 + \frac{1}{24} Z_\sym x^4+\dots \, , 
\end{eqnarray}
where the density expansion parameter is defined as $x=(n-n_\sat)/(3n_\sat)$, with $n_\sat$ being the saturation density.

In the MM, the specific density and isospin asymmetry dependence of the kinetic term is preserved, see Refs.~\cite{Margueron2018a,Somasundaram2021} for more details. One can then decide to fix the NEP to the known empirical values and use the unknown ones to explore extrapolations at high density~\cite{Margueron2018b}, or one can also use the MM as a fit of actual predictions. In the present study, we will employ this second application of the MM. We thus adjust the MM to the BSk24 Skyrme predictions by fixing the low order NEP to be equal to those of BSk24~\cite{Goriely2013} ($E_\sat$/$E_\sym$, $n_\sat$, $L_{\sym}$, $K_\sat$/$K_\sym$), the effective mass ($m^*_\sat$) and the splitting of the effective mass ($\Delta m^*$) and varying the unknown high order ones ($Q_\sat$/$Q_\sym$, $Z_\sat$/$Z_\sym$) and the low density parameters ($b_\sat$/$b_\sym$) in a density range from zero up to 10 $n_\sat$. The parameters resulting from the fit in uniform matter are given in Tab.~\ref{tab:bsk24}.

\begin{center}
\begin{table}[tb]
\centering
\caption{Parameters of the MM adjusted to BSk24 predictions in SM and NM. We show first the NEP $E_\sat$, $E_\sym$, $n_\sat$, $L_\sym$, $K_\sat$, $K_\sym$, $m^*_\sat$ and $\Delta m^*$ which are taken from Ref.~\cite{Goriely2013}. The second line shows the parameters $Q_\sat$, $Q_\sym$, $Z_\sat$, $Z_\sym$, $b_\sat$ and $b_\sym$ which are deduced from the fit to BSk24 pressure in SM and NM.}
\label{tab:bsk24}
\begin{tabular}{@{}l*{15}{l}}
\br
NEP & $E_\sat$ & $E_\sym$ & $n_\sat$ & $L_\sym$ & $K_\sat$ & $K_\sym$ & $m^*_\sat$ & $\Delta m^*$  \\
\mr
& -16.048 & 30.0 & 0.1578 & 46.3967 & 245.52 & -37.60 & 0.80 & 0.21 \\
\br
Fit & $Q_\sat$ & $Q_\sym$ & $Z_\sat$ & $Z_\sym$ & $b_\sat$ & $b_\sym$ \\
\mr
& -49.22 & 361.70 & -910.69 & -932.13 & 1.28 & 0.02\\
\br
\end{tabular}
\end{table}
\end{center}

\begin{figure}[h]
\includegraphics[width=21pc]{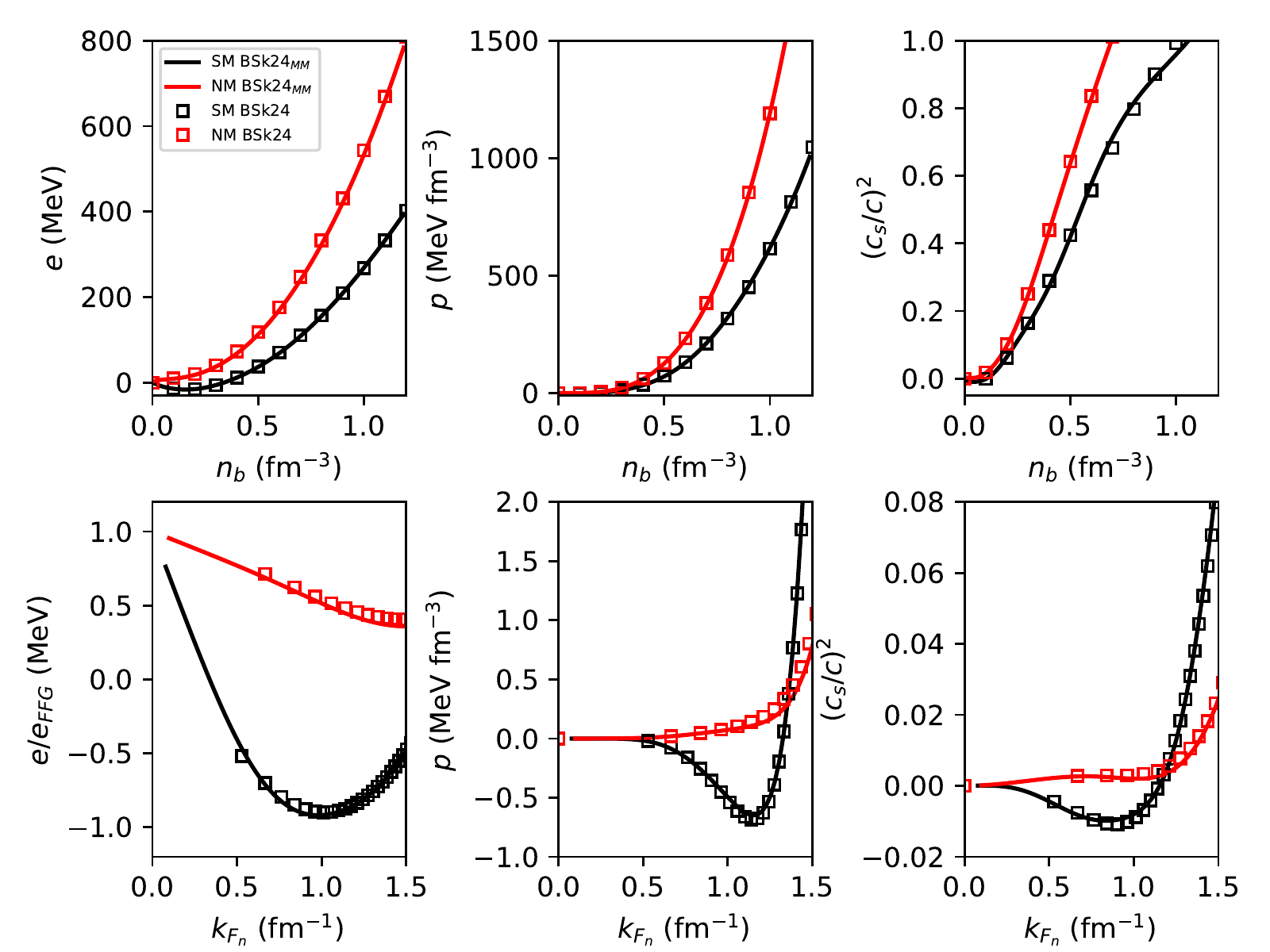}\hspace{2pc}%
\begin{minipage}[b]{14pc}\caption{\label{fig:unifbsk24}Energy per particle (left), pressure (center), and sound speed (right) from BSk24 (squares) and the MM (curves) in SM and NM and as function of the density (top panels) or as function of the neutron Fermi momentum $k_{F_n}$ (bottom panel). The free Fermi gas energy per particle $e_{FFG}$ is employed to scale the energy per particle $e$.}
\end{minipage}
\end{figure}

We show in Fig.~\ref{fig:unifbsk24} the good agreement of the MM fitted to the original BSk24 Skyrme predictions in uniform matter.
The three top panels show the energy per particle $e$, the pressure $p$ and the sound speed square $(c_s/c)^2$ as function of the density in the whole interval considered by the fit. The bottom panel shows the same quantities zoomed at the low densities of relevance in the crust of NS. Fig.~\ref{fig:unifbsk24} shows that a single parameterisation of the MM is able to mimic the BSk24 predictions from the lowest to the highest densities. This ability was also illustrated for the $\chi_{EFT}$ Hamiltonian predictions in Ref.~\cite{Somasundaram2021}.

\section{Neutron star crust properties from BSk24 Skyrme interaction}

In this section, we analyze the predictions for the NS crust obtained from the CLDM based on the MM adjusted to BSk24 Skyrme interaction in uniform matter. We also compare such predictions with the ones of more refined approaches: ETF and ETFSI employing the same BSk24 Skyrme interaction.

The CLDM model used here is the FS4 described in \cite{Grams2022b}, which contains Coulomb direct and exchange terms, surface and curvature energies. We optimize the CLDM by fitting to all experimental masses from the 2012 atomic masses evaluation (AME2012) \cite{Wang_2012}.
Details of the CLDM model and fit to the nuclear masses are well detailed in our recent work \cite{Grams2022a}. We briefly remind here the expression for the surface energy.
The surface energy is proportional to the surface tension $\sigma_\surf(I_\cl)$, where $I_\cl$ is the cluster asymmetry $I_\cl = (N_\cl - Z_\cl)/A_\cl$, and scales as $A_\cl^{2/3}$, where $A_\cl$ is the mass number in the cluster. It reads
\begin{equation}
   E_\surf(A_\cl,I_\cl,n_{\cl}) = 4 \pi R^2_\cl \sigma_\surf(I_\cl) 
   =  4 \pi r^2_{\cl}  \sigma_\surf(I_\cl) A_\cl^{2/3} \, ,
   \label{eq:esurf}
\end{equation}
with $R_\cl=r_{cl}(n_\cl)A_\cl^{1/3}$, $r_{cl}^3(n_\cl)=3/(4 \pi n_{\cl})$, with $n_\cl$ being the cluster density, and $\sigma_\surf(I_\cl)$ is expressed, as suggested in \cite{LattimerSwesty1991}, as
\begin{equation}
\sigma_\surf(I_\cl)=\sigma_{\surf, \sat} \frac{2^{p_\surf+1}+b_\surf}{Y_p^{-p_\surf}+b_\surf+(1-Y_p)^{-p_\surf}}, 
\label{eq:sigma}
\end{equation}
where $Y_p = Z_\cl/A_\cl = (1-I_\cl)/2$ is the cluster proton fraction and $\sigma_{\surf, \sat}$ is a parameter that determines the surface tension in symmetric nuclei. 
The parameter $p_\surf$ plays an important role at large isospin asymmetries. It is usually fixed to be $p_\surf=3$ since the seminal contribution~\cite{LattimerSwesty1991}, but a small variation around $3$ plays an important role at large asymmetries, which occurs around the core-crust transition densities in NSs~\cite{Carreau2019b,Grams2022a}. In Ref.~\cite{Grams2022a}
we varied it in the range [2.5, 3.0, 3.5] to demonstrate how important is this parameter at the edge of the inner crust, where clusters are very neutron rich. This small variation on $p_{\surf}$ can modify the crust-core transition and the composition of the inner crust, as shown in Fig 10 of Ref.~\cite{Grams2022a}. 
Since it influences high asymmetries, $I_\cl \apprge 0.3$, see for example Fig. 4b of \cite{Grams2022a}, this parameter is not constrained by nuclear masses. In this paper, we explore the impact of changing $p_\surf$ in the range [2.9, 3.1, 3.3, 3.5]. The figures for the CLDM-MM-BSk24 equation of state (EOS) are therefore shown within a band generated by varying $p_\surf$ inside this range.
The coefficients $\sigma_{\surf, \sat}$, $\sigma_{\surf, \sym}$, $\sigma_{\curv}$ and $\beta_\curv$ obtained after the fit to the nuclear masses are shown in Tab.~\ref{jfonts}.
These fits are obtained by varying the FS model~\cite{Grams2022a} (going from FS1 to FS4) as well as the surface parameter $p_\surf$. The loss function $\chi_E$ measures the standard deviation between the experimental masses and the model predictions.

The CLDM is expressed in the so-called coordinate or r-representation \cite{Papakonstantinou2013} of the nuclear cluster considering that the neutron fluid does not penetrate inside the volume occupied by the cluster, while the ETF and ETFSI are expressed in the energy-representation or e-representation, where the neutron fluid overlaps the nuclear clusters. The r-representation treats the nuclear clusters as classical hard spheres while the e-representation is closer to the quantum nature of the system. These two representations of the density distribution of the particles in the system are however approximate ways to describe the real quantum nature of the system based on wave functions. We refer to Ref.~\cite{Papakonstantinou2013} for more details on the comparison of these two representations with the quantum description of dilute clusters in the crust of NSs.

In the r-representation, the total baryon density $n_B$ inside the Wigner-Seitz (WS) cell reads
\begin{equation}
n_B = \frac{A_{\cl}+N_{n,fluid}}{V_{\WS}} = n_{cl} u + n_{n,fluid} (1-u) \, ,
\end{equation}
where $N_{n,fluid}$ is the number of neutrons in the external fluid.
We introduced the volume fraction $u$ defined as $u=V_{\cl}/V_{\WS}$, with $V_\cl$ ($V_\WS$) being the cluster (Wigner-Seitz cell) volume, the cluster density $n_\cl = A_{\cl}/V_{\cl}$ and the neutron fluid density $n_{n,fluid} = N_{n,fluid} /(V_\WS - V_\cl )$. 

\begin{center}
\begin{table}[tb]
\centering
\caption{\label{jfonts}Parameters for the FS terms and $\chi_{E}$  of the present model obtained optimizing the MM-CLDM BSk24 on the nuclear chart AME2012 \cite{Wang_2012}. 
} 
\begin{tabular}{@{}l*{15}{l}}
\br
FS model & $ p_\surf$ & $ C_\coul$ & $\sigma_{\surf,\sat}$  & $\sigma_{\surf,\sym}$  & $\sigma_\curv$  & $\beta_{\curv}$  & $\chi_{E}$\\
\mr
FS1 & 3.1  & 0.966 & 1.137 &  1.421  &   0.0 &   0.0 & 3.27 \\
FS2 & 3.1  & 0.955 & 1.159 &  1.163  &   0.0 &   0.0 & 2.92 \\
FS3 & 3.1  & 0.966 & 1.113 &  1.792  &   0.104 &   0.614 & 2.73\\
\br
FS4 & 2.9  & 0.975 & 1.185 &  1.984  &   0.069 &   0.551 & 2.68\\
FS4 & 3.1  & 0.975 & 1.184 &  1.955  &   0.068 &   0.557 & 2.67\\
FS4 & 3.3  & 0.975 & 1.183 &  1.925  &   0.067 &   0.563 & 2.66\\
FS4 & 3.5  & 0.975 & 1.181 &  1.895  &   0.066 &   0.570 & 2.66\\
\br
\end{tabular}
\end{table}
\end{center}


The ETF and ETFSI EOS from Ref. \cite{Pearson2018} are computed in the e-representation of the WS cell, where a background of neutrons and protons occupy the whole WS cell.
The baryon density in the e-representation reads,
\begin{equation}
n_B= (n_{\cl} - n_{n,fluid}) u + n_{n,fluid} \, ,
\end{equation}
where we use the conservation of particle number in the nuclear cluster,
\begin{equation}
A_{r-rep}^{\cl}=n_{\cl} V_{\cl} = A_{e-rep}^{\cl}+ n_{n,fluid} V_{\cl}.
    \label{relation.r-e}
\end{equation}{}


\begin{figure}[h]
\includegraphics[width=22.2pc]{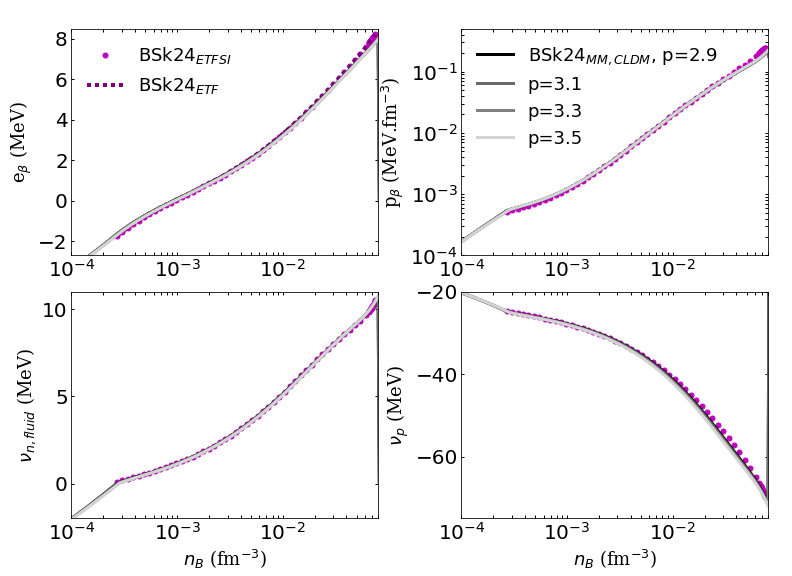}\hspace{2pc}%
\begin{minipage}[b]{14pc}\caption{\label{fig:thermo}Energy per particle (top left), pressure (top right), neutron chemical potential (bottom left) and proton chemical potential (bottom right).  Comparison between the MM CLDM version of BSk24 and the original ETF and ETFSI model.}
\end{minipage}
\end{figure}

In the following comparisons, we use the relations between the neutron numbers in the two representations: 
$N^{\cl}_{ETF} = N^\cl_{CLDM} - n_{n,fluid} V_{\cl}$, where $V_\cl$ is defined from $R_\cl$ given from the CLDM approach. Note that the prescription we employ to relate the CLDM and ETFSI results is approximate. For instance, the cluster volume is not exactly the same in both models, especially since the ETFSI treats differently the distributions of neutrons and protons, while the CLDM does not. This small discrepancy however does not strongly impact the results.


We show in Fig. \ref{fig:thermo} the following thermodynamic quantities at $\beta$-equilibrium: the energy per particle $e_\beta$ (top left), the pressure $p_\beta$ (top right), the neutron fluid chemical potential $\nu_{n,fluid}$ and the proton
chemical potential $\nu_p$ 
(bottom right). Note that we do not represent $\nu_{p,fluid}$ since there is no proton fluid in the CLDM. Fig. \ref{fig:thermo} shows that there is a very good agreement of the CLDM with the ETFSI predictions for all these quantities. It reflects that the thermodynamics of the EOS is strongly constrained by the bulk properties.


\begin{figure}[h]
\includegraphics[width=22.2pc]{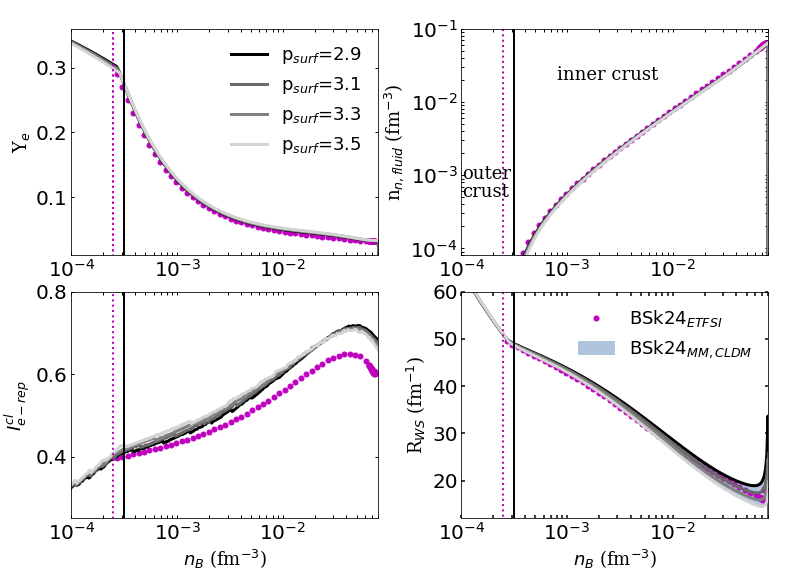}\hspace{2pc}%
\begin{minipage}[b]{14pc}\caption{\label{fig:asymmetry}Electron fraction $Y_e$ (top left), neutron fluid density (top right), cluster asymmetry $I_\cl$ (bottom left) and Wigner-Seitz cell radius $R_\WS$ (bottom right) as functions of the baryon density. The vertical lines depict the boundary between the outer and inner crust for each model.}
\end{minipage}
\end{figure}

We plot in Fig. \ref{fig:asymmetry} the electron fraction $Y_e$, neutron fluid density, cluster asymmetry $I_\cl$ and Wigner-Seitz cell radius $R_\WS$ as functions of the baryon density. We show the cluster asymmetry considering the e-representation ($I_\cl^{e-rep}$). Magenta dots represent ETFSI calculations, continuous lines in gray scale and within a blue band show the CLDM results for different values of $p_\surf$. Note that the CLDM-MM calculations are in very good agreement with the ETFSI ones with the exception of the cluster asymmetry $I_\cl$. At the transition between the outer and inner crusts, the difference between these two theoretical models is minimal, and it increases as function of the total density inside the inner crust. It reflects the influence of the increasing amount of neutron fluid in the nuclear clusters as the density increases, which impacts the FS terms: The two main differences come from the existence of a neutron skin and  proton shell effects in the ETFSI approach which are missing in the CLDM.
Note also the very small influence of the parameter $p_\surf$ on these quantities, except for $R_{WS}$ in the densest region of the crust. This region coincides however with the one of the pasta phase, which is not described by our models. It will therefore be interesting in the future to improve the CLDM-MM for the description of the densest region of the inner crust.

\begin{figure}[tb]
\centering
\includegraphics[scale=0.34]{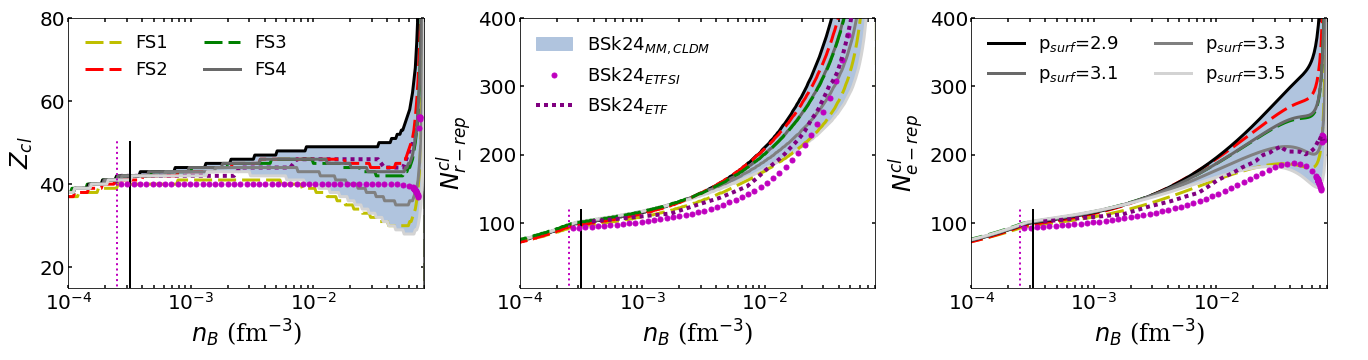}
\caption{Crust composition $Z_\cl$, $N_\cl$ as function of the baryon density. The number of neutrons is calculated in the r-representation ($N_\cl^{r-rep}$ middle panel) and in the e-representation ($N_\cl^{e-rep}$ right panel). We explore the impact of the leptodermous expansion by showing the results for FS1-FS4 in the CLDM, the impact of $p_\surf$ by varying $p_\surf$ from 2.9 to 3.5 in the CLDM, and finally, we compare CLDM with EFT and EFTSI predictions. The blue band correspond to the CLDM uncertainty in the parameter $p_\surf$.}
\label{fig:NZ}
\end{figure}

To study in more details the origin of the difference between the modelings on $I_\cl$, we show in Fig.~\ref{fig:NZ} a comparison of $Z_\cl$ and $N_\cl$ as function of $n_{B}$. The number of protons in the nuclear cluster $Z_\cl$ is shown in the left panel and neutron number in the nuclear cluster $N_\cl$ in the two other panels (r-representation in the middle panel and e-representation in the right panel). The blue band contains CLDM-MM predictions where the parameter $p_\surf$ is varied, and the different dashed lines show the impact of the FS description ordered within the leptodermous expansion. The dotted dark purple line shows the ETF results. It is interesting to remark that by fixing $p_\surf\approx 3.1$ there is a very good overlap between the CLDM-MM and the ETF predictions for $Z_\cl$. The same feature exists for $N_\cl$, but for $p_\surf\approx 3.3$. This small discrepancy may originate from the neutron skin, which is not included in the CLDM-MM. The impact of the neutron skin is however small compared to the FS terms of higher rank in the leptodermous expansion. The convergence of the leptodermous expansion is then still satisfied for the neutron skin.

The shell effects shown in Fig.~\ref{fig:NZ} are seen as the difference between ETF and ETFSI predictions for $Z_\cl$ and $N_\cl$. The proton shell effects stabilises $Z_\cl$ to proton shell or sub-shell closure in almost all the inner crust ($Z_\cl=40$ for BSk24). It is only in the densest region of the inner crust, the so-called pasta phase, that $Z_\cl$ departs from 40. Since shell or sub-shell closure occurs at fixed proton numbers, expected to be 20, 28, 40, 50 and 82 from nuclear phenomenology, the impact of the shell effect could be larger on $Z_\cl$ and $N_\cl$ than the effect of FS4 or of the neutron skin. For BSk24, since CLDM and ETF predict values for $Z_\cl$ slightly above 40 (of the order of 42-44), the stabilization from the proton shell effects in ETFSI decreases $Z_\cl$ down to 40. As a consequence, $N_\cl$ is also reduced down to approximately conserve the isospin asymmetry. The shell effects however have a very small contribution to the thermodynamical quantities, as shown in Fig.~\ref{fig:thermo} for instance. So despite its large impact in relative value for $Z_\cl$ and $N_\cl$, shell effects are still small for the thermodynamical quantities, as expected from the leptodermous expansion.

\section{Conclusions}

In this study, by fixing the nuclear interaction to be the Skyrme BSk24 force, we have shown a good convergence between CLDM, ETF and ETFSI predictions for the thermodynamical properties and we have analyzed the origin of the differences observed for the nuclear cluster composition ($Z_\cl$, $N_\cl$). The main features of the present work are listed below:
\begin{enumerate}
\item A unified EOS using the CLDM based on the BSk24 MM is constructed. 
\item The analysis of the FS terms on the NS crust has shown that the fit to the nuclear masses is a good strategy to optimize the CLDM parameters; this approach however does not constrain accurately the surface parameter $p_\surf$, which is varied here in the following range [2.9, 3.1, 3.3, 3.5].
\item The impact of various improvements in the FS terms has been analyzed: from FS1 to FS4 in CLDM-MM~\cite{Grams2022a}, ETF (neutron skin) and ETFSI (shell effects)~\cite{Pearson2018}.
\item The impact of the FS terms scales according to the leptodermous expansion for the thermodynamical quantities, as anticipated in Ref.~\cite{Grams2022a}.
\item Neutron skin moderately impacts the cluster configuration ($Z_\cl$, $N_\cl$), while proton shell effects have a large impact on stabilizing the proton number $Z_\cl$ to shell or sub-shell closure. 
\item The parameter $p_\surf$ has still a large impact in the CLDM-MM prediction for the densest region of the inner crust, as already shown in \cite{Carreau2019b, Grams2022a}. The optimal value to reproduce $Z_\cl$ and $N_\cl$ from ETF calculations based on BSk24 Skyrme interaction~\cite{Pearson2018} is $p_\surf=3.1$-$3.3$, where the difference reflects to absence of neutron skin in the present CLDM-MM.
\end{enumerate}

In conclusion, the CLDM-MM furnishes a very good tool to compute the thermodynamical properties of the crust and to construct core-crust unified EOS. Some improvements are however still possible, in particular to better describe the nuclear cluster configurations ($Z_\cl$ and $N_\cl$). Our future plans to develop the CLDM-MM are thus to implemented the contribution from neutron skin, as suggested in Refs.~\cite{DouchinHaensel2000,Vinas2017} as well as the proton shell effects, which are less easy to implement. A simple model was however suggested in Ref.~\cite{Tews2017}. One has however to keep in mind that the largest source of uncertainties for the NS crust properties originates from the unknown nuclear interaction. The largest corrections to the present results are obtained by changing one nuclear model to another, see for instance Refs.~\cite{Pearson2018,Grams2022a,Grams2022b}. Reducing these uncertainties requires complementary constraints from both nuclear physics laboratories and NS observations. It is therefore important to confront unified models for the NS EoS to observation data, as well as to constrain them to theoretical expectations, such as for instance the ones suggested by $\chi$EFT in low density NM~\cite{Tews2017,Grams2021a,Grams2022b}.

\ack
G.G., J.M. and R.S. thank Sanjay Reddy for fruitful discussions regarding modeling of NS crust properties.
G.G., J.M. and R.S. are supported by CNRS grant PICS-08294 VIPER (Nuclear Physics for Violent Phenomena in the Universe), the CNRS IEA-303083 BEOS project, the CNRS/IN2P3 NewMAC project, and benefit from PHAROS COST Action CA16214. This work is supported by the LABEX Lyon Institute of Origins (ANR-10-LABX-0066) of the \textsl{Universit\'e de Lyon} for its financial support within the program \textsl{Investissements d'Avenir} (ANR-11-IDEX-0007) of the French government operated by the National Research Agency (ANR), Fonds de la Recherche Scientifique-FNRS (Belgium) under Grant Number IISN 4.4502.19. NC and SG are F.R.S.-FNRS senior research associates.

\bibliographystyle{iopart-num}       
\bibliography{ref.bib}   

\end{document}